\documentstyle[pra,aps,multicol,epsfig]{revtex}

\newcommand{\eqref}[1]{Eq.~(\ref{#1})}

\begin{document}

\title{Solitons in nonlocal nonlinear media: exact results}

\author{Wies{\l}aw Kr\'olikowski}

\address{Australian Photonics Cooperative Research Centre,
Laser Physics Centre, \\
Research School of Physical Science and Engineering, \\
The Australian National University, Canberra ACT 0200, Australia}

\author{Ole Bang}

\address{Department of Mathematical Modelling, \\
Technical University Denmark, Building 305/321,
DK-2800 Kgs.~Lyngby, Denmark}

\maketitle

\begin{abstract}
We investigate the propagation of one-dimensional bright and dark
spatial solitons in a nonlocal Kerr-like media, in which the 
nonlocality is of general form.  
We find an exact analytical solution to the nonlinear propagation 
equation in the case of weak nonlocality.  
We study the properties of these solitons and show their stability.
\end{abstract}


\begin{multicols}{2}

\narrowtext

\section{Introduction}

Let us consider a phenomenological model of nonlocal nonlinear Kerr 
type media, in which the refractive index change $\Delta n$ induced by
a beam with intensity $I(x,z)$ can be represented in general form as
\begin{equation}
  \label{nonlocal}
  \Delta n(I)= \pm\int_{-\infty}^{\infty}R(x'-x)I(x',z)dx',
\end{equation}
where the positive (negative) sign corresponds to a focusing 
(defocusing) nonlinearity and {\em x} and {\em z} denote
transverse and propagation coordinates, respectively. The real, 
localized, and symmetric function $R(x)$ is the response function of 
the nonlocal medium, whose width determines the degree of nonlocality.
For a singular response, $R(x)=\delta(x)$, the refractive index change
becomes a local function of the light intensity, $\Delta n(I)=\pm
I(x,z)$, i.e.~the refractive index change at a given point is solely
determined by the light intensity at that very point.  
With increasing width of $R(x)$ the light intensity in the vicinity 
of the point $x$ also contributes to the index change at that point. 
In the limit of a highly nonlocal response Snyder and Mitchell showed 
that the beam evolution was described by the simple equation for a 
linear harmonic oscillator \cite{Snyder97}. 
The influence of nonlocality of the nonlinear response on the dynamics
of beams was illustrated for the special logarithmic nonlinearity, which
allows exact analytical treatment \cite{Snyder99}.

While Eq.~(\ref{nonlocal}) is a phenomenological model, it nevertheless 
describes several real physical situations. 
Possible physical mechanisms responsible for this type of nonlinear 
response includes various transport effects, such as heat conduction in 
materials with thermal nonlinearity \cite{thermal1,thermal2,thermal3}, 
diffusion of molecules or atoms accompanying nonlinear light propagation 
in atomic vapours \cite{suter}, and drift and/or diffusion of photoexcited
charges in photorefractive materials \cite{pr1,pr2}. 
A highly nonlocal nonlinearity of the form (\ref{nonlocal}) has also been 
identified in plasmas \cite{cusp,litvak,juul,df} and it appears as a result 
of many body interaction processes in the description of Bose-Einstein 
condensates \cite{bose}.

Even though it is quite apparent in some physical situations that 
the nonlinear response in general is nonlocal (as in the case of thermal 
lensing), the nonlocal contribution to the refractive index change was 
often neglected \cite{dark1,dark2}.  
This is justified if the spatial scale of the beam is large compared 
to the characteristic response length of the medium (given by the 
width of the response function). However, for very narrow beams the 
nonlocality can be of crucial importance and has to be taken into account.
For instance, it has been shown theoretically that a weak nonlocal 
contribution arrests collapse (catastrophic self-focusing) of high 
power optical beams in a self-focusing medium and leads to the 
formation of stable 2D (diffracting in two transverse dimensions) 
solitons \cite{juul,df,turitsyn,abe98}.  
On the other hand, a purely nonlocal nonlinearity leads to formation of 
so-called cusp solitons, which, however, are unstable \cite{cusp}. 

Some of the consequences of nonlocality in the nonlinear response have been 
observed experimentally. Suter and Blasberg reported stabilization of 2D
solitary beams in atomic vapors due to atomic diffusion, which carries 
excitation away from the interaction region \cite{suter}. 
Also, the discrepancy between the theoretical model of dark solitons and 
that observed experimentally in a medium with thermal nonlinearity has 
been associated with nonlocality of the nonlinearity \cite{dark1,dark2}.

Here, we investigate the propagation of 1D beams in nonlinear media 
having a weakly nonlocal response of the general form (\ref{nonlocal}). 
Our goal is to find exact analytical solutions for bright and dark spatial 
solitons and use them to determine soliton properties, such as existence
and stability.  
We start with the paraxial wave equation describing propagation of a 1D
beam with envelope function $\psi=\psi(x,z)$ and intensity 
$I=I(x,z)=|\psi(x,z)|^2$,
\begin{equation}
  \label{nls1}
  i\partial_z\psi + \frac{1}{2}\partial_x^2\psi + \Delta n(I) \psi = 0.
\end{equation}
When the nonlocality is weak, i.e.~when the response function $R(x)$ is
narrow compared to the extent of the beam, we can expand $I(x',z)$ around
the point $x'=x$ to obtain
\begin{equation}
  \label{laplacian}
  \Delta n(I) = \pm ( I + \gamma\partial_x^2I ),
\end{equation}
where the nonlocality parameter $\gamma>0$ is given by
\begin{equation}
  \gamma=\frac{1}{2}\int_{-\infty}^{\infty}R(x)x^2dx,
\end{equation}
and where we have assumed that the response function is normalized,
$\int_{-\infty}^{\infty}R(x)dx=1$.
Note that for $R(x)=\delta(x)$, $\gamma=0$ and Eq.~(\ref{laplacian})
describes the local Kerr nonlinearity.
For weakly nonlocal meda $\gamma\ll1$ is a small parameter.

\begin{figure}
  \epsfig{file=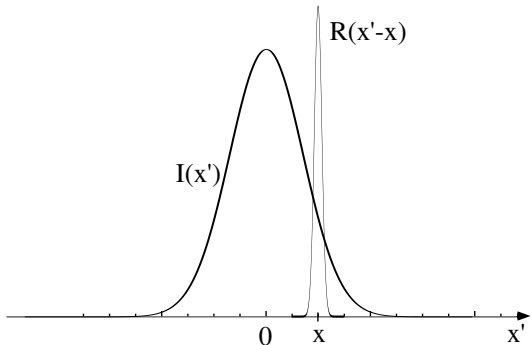,width=7cm,angle=0}
  \caption{Intensity profile $I(x')=I(x',z)$ and response 
  function $R(x'-x)$ in the weakly nonlocal limit.} 
  \label{fig1}
\end{figure}

Substituting $\Delta n(I)$, given by Eq.~(\ref{laplacian}), into 
Eq.~(\ref{nls1}) gives the modified nonlinear Schr\"{o}dinger equation
\begin{equation}
  \label{NLS1}
  i\partial_z\psi + \frac{1}{2}\partial_x^2\psi \pm 
  (|\psi|^2 + \gamma\partial_x^2|\psi|^2) \psi = 0.
\end{equation}
The weak nonlocality appears thus as a perturbation to the local 
nonlinear refractive index change. For a single peak beam in a
self-focusing medium this perturbation is of negative sign in the 
central part of the beam, where it serves to decrease the refractive 
index change. Hence, even for very narrow and sharp intensity
distributions, the resulting self-induced waveguide will be wide and a
smooth function of the transverse coordinates. In some sense, this is
similar to saturation of the nonlinearity. One may
therefore expect that certain features of solitons of Eq.~(\ref{NLS1}) 
will be reminiscent of those exhibited by solitons in saturable nonlinear 
media. It transpires, however, that nonlocality also results in new effects, 
especially for self-defocusing nonlinearities.
In the following we consider separately the case of self-focusing and 
self-defocusing nonlinearities.

\section{Bright solitons}

For a self-focusing nonlinearity the sign of the refractive index change 
is positive. We search for a stationary bright soliton solution to 
Eq.~(\ref{NLS1}) of the form
\begin{equation}
  \psi(x,z) = u(x) \exp(i\Gamma z),
\end{equation}
where the profile $u=u(x)$ is real, symmetric, and exponentially 
localized and the propagation constant $\Gamma>0$ is positive.
For this solution Eq.~(\ref{NLS1}) reduces to
\begin{equation}
  \partial_x^2u + 2(u^2-\Gamma)u + 2\gamma u\partial_x^2u^2 = 0,
\end{equation}
which can be integrated once to give the equation
\begin{equation}
  \label{stationary_eq}
  (1+4\gamma u^2)(\partial_xu)^2 + (u^2-2\Gamma)u^2 = C,
\end{equation}
where $C$ is an integration constant. For exponentially localized solutions 
$C=0$. Using that $u(x)$ has its maximum $u_0$ at the center $x=0$ we futher 
obtain the well-known relation between the propagation constant $\Gamma$ 
and the amplitude $u_0$
\begin{equation}
  u_0^2 = 2\Gamma .
\end{equation}
Equation (\ref{stationary_eq}) can therefore be simplified to
\begin{equation}
  \label{deriv}
  (\partial_xu)^2 = u^2(u_0^2-u^2)/(1+4\gamma u^2).
\end{equation}
Interestingly, had the local and nonlocal contributions been of opposite 
signs, i.e.~$\gamma<0$ (as can happen for intense laser beams in plasmas), 
then a solution to Eq.~(\ref{deriv}) would only exist if the peak 
intensity $\rho_0=u_0^2$ is less than the critical value 
$\rho_{\rm cr}=1/|4\gamma|$ (see also \cite{litvak}).
A final integration leads to
\begin{equation}
  \label{solution1}
  \pm x = \frac{1}{u_0}\tanh^{-1}\left(\frac{\sigma}{u_0}\right) + 
          \sqrt{4\gamma}\tan^{-1}(\sqrt{4\gamma}\sigma)
\end{equation}
where we have defined the intensity $\rho=u^2$ and the normalized 
intensity $\sigma^2=(\rho_0-\rho)/(1+4\gamma\rho)$.
This implicit relation gives the profile of bright spatial solitons 
propagating in weakly nonlocal Kerr-like media. In the local limit, 
$\gamma=0$, we recover from Eq.~(\ref{solution1}) the well-known 
profile $u(x)=u_0{\rm sech}(u_0x)$ of the 1D bright soliton in a 
Kerr medium.

\begin{figure}
  \epsfig{file=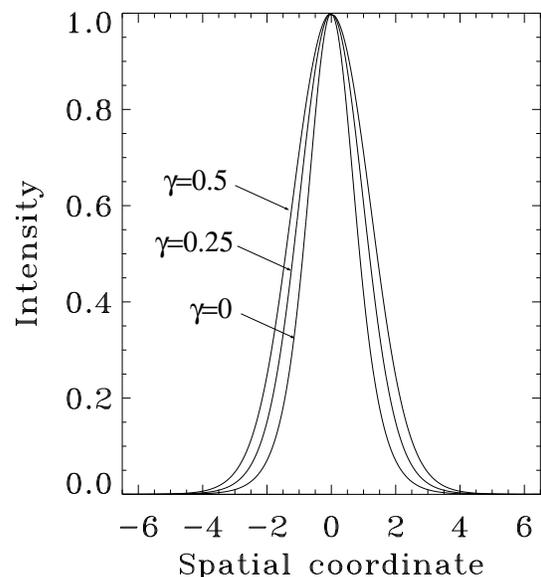,width=7cm,angle=0}
  \caption{Intensity profiles of bright solitons with unit amplitudes
  $\rho_0=1$ for different degrees of nonlocality $\gamma$.} 
  \label{fig2}
\end{figure}

In Fig.~\ref{fig2} we show the intensity profile of the solution 
(\ref{solution1}) for different values of the nonlocality parameter 
$\gamma$. 
Evidently, an increase in the degree of nonlocality results in an 
increase of the soliton width - nonlocality smooths out the refractive 
index profile thereby leading to a broadening of the beam. 
This effect is more clearly seen in Fig.~3 where we plot the width of
the bright solitons (FWHM of the intensity profile) versus the degree
of nonlocality $\gamma$ for different peak intensities; the soliton 
width increases monotonically with the degree of nonlocality.

\begin{figure}
  \epsfig{file=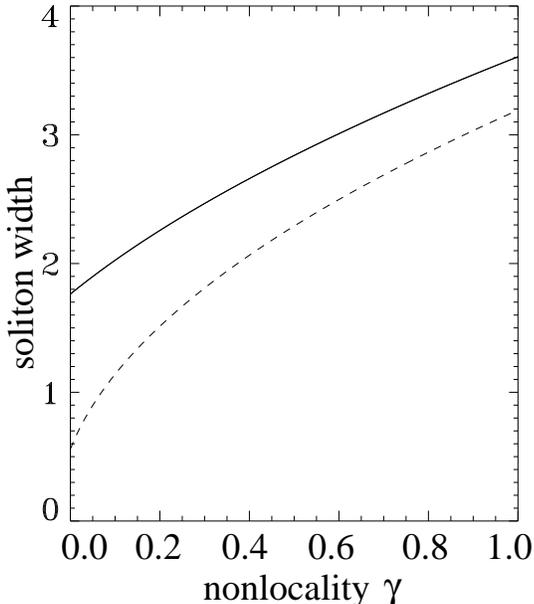,width=7cm,angle=0}
  \caption{Soliton width versus degree of nonlocality $\gamma$ 
  for peak intensities $\rho_0$=1 (solid) and $\rho_0$=10 (dashed).} 
  \label{fig3}
\end{figure}

An important aspect of any family of soliton solutions is their
stability properties. For single peak solitons stability is
determined by the dependence of the power 
\begin{equation}
  P = \int_{-\infty}^\infty |\psi(x,z)|^2dx,
\end{equation} 
on the propagation constant $\Gamma$.  
Solitons are stable if $dP/d\Gamma>0$ and unstable otherwise \cite{ladtke}.  
For the model considered here $P(\Gamma)$ can be found analytically to
\begin{equation}
  \label{Pgen}
  P = \sqrt{\rho_0} + \frac{1+4\gamma\rho_0}{\sqrt{4\gamma}} 
  \tan^{-1}(\sqrt{4\gamma\rho_0}),
\end{equation}
by using Eq.~(\ref{deriv}). For weakly nonlocal media with $\gamma\ll1$
the power is approximately given by
\begin{equation}
  \label{Pweak}
  P = P_0(1+\frac{4}{3}\gamma\rho_0-\frac{16}{15}\gamma^2\rho_0^2+...),
\end{equation}
where $P_0=2\sqrt{\rho_0}$ is the soliton power in the limit of a local
response with $\gamma$=0.
The derivative $dP/d\Gamma$ can easily be found from Eq.~(\ref{Pgen}) 
and it transpires that the power is a monotonically increasing function 
of the propagation constant (for $\gamma>0$), indicating that the 
solitons are stable. 

Interestingly again, had $\gamma$ been negative, then the bright solitons 
would exist for sufficiently low peak intensities, $\rho_0<1/|4\gamma|$, 
but be unstable for $\rho_0>0.7/|4\gamma|$, for which $dP/d\Gamma$ is 
negative (see again \cite{litvak}).

To demonstrate the stability of the bright solitons for $\gamma>0$ we 
numerically integrated Eq.~(\ref{NLS1}) using the split-step fourier 
method and the exact soliton solution as initial condition. 
In all simulations (for both bright and dark solitons) we used a 
steplength of $dz=0.001$ and a transverse resolution of $dx=0.05$.
Results of the numerical simulations are shown in Fig.~4 where we 
demonstrate propagation and collision of two bright solitons with
unit amplitude $\rho_0=1$. 
It is evident that the solitons propagate in a stable manner. The 
collision, on the other hand, is not completely elastic and causes 
the soliton amplitude and width to oscillate slightly.

\begin{figure}
  \epsfig{file=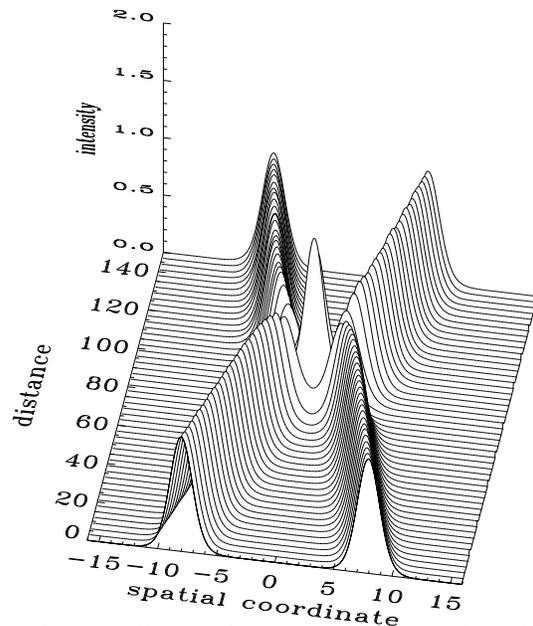,width=7cm,angle=0}
  \caption{Collision of unit amplitude bright solitons
  in a weakly nonlocal Kerr-like medium with $\gamma=0.1$} 
  \label{fig4}
\end{figure}

The bright solitons of the nonlocal nonlinear Schr\"odinger Eq.~(\ref{NLS1}) 
were considered by Davydova and Fischuk in the context of focusing of upper
hybrid waves in plasma \cite{df}. In particular, the existence of bright 
soliton solutions and their stability was reported by these authors. 
However, the explicit expression (\ref{solution1}) was not given and
no numerical confirmation was presented.

\section{Dark solitons}

We now consider the impact of weak nonlocality in the case of a
self-defocusing nonlinearity, which corresponds to a negative sign 
in Eq.(\ref{laplacian}). We introduce a new spatial variable
$\zeta=x-Vz$, with $V$ being the soliton transverse velocity, 
and look for stationary solutions of the form 
\begin{equation}
  \psi(x,z) = \sqrt{\rho(\zeta)}\exp(i\Gamma z + i\phi(\zeta)),
\end{equation} 
where $\rho(\zeta)$ is the soliton intensity and $\phi(\zeta)$ its
phase. Substitution of this form into Eq.~(\ref{NLS1}) yields
\begin{eqnarray}
  \label{dark2}
  & & \partial_\zeta\{\rho(\partial_\zeta\phi-V)\} = 0,\\
  \nonumber
  & & 2\rho(1-4\gamma\rho)\partial_\zeta^2\rho - (\partial_\zeta\rho)^2
  - 4\rho^2(\partial_\zeta\phi)^2 + 8V\rho^2\partial_\zeta\phi  \nonumber \\
  \label{dark1}
  & & - 8\rho^2(\Gamma+\rho) = 0 .
\end{eqnarray}
We are interested in dark solitons, i.e.~solutions with an intensity 
dip on a constant background.  
Integrating the system (\ref{dark2}-\ref{dark1}) once we obtain that the
soliton background intensity $\rho_0=u_0^2$ determines the soliton propagation 
constant $\Gamma$ and that the center intensity $\rho_1=u_1^2$ determines the
transverse velocity $V$, through the relations 
\begin{equation}
  V^2 = \rho_1 , \;\;\; \Gamma = -\rho_0,
\end{equation}
which are exactly the same as those obtained for a purely local response.  
We further find that the soliton intensity $\rho(\zeta)$ is governed by 
the equation
\begin{equation}
  \label{dark-ODE}
  (\partial_\zeta\rho)^2 = 4(\rho-\rho_1)(\rho_0-\rho)^2/(1-4\gamma\rho).
\end{equation}
Obviously, a solution to Eq.~(\ref{dark-ODE}) exists only if the
background intensity does not exceed a certain critical value,
\begin{equation}
  \label{critical}
   \rho_0 < \rho_{\rm cr} = 1/|4\gamma|,
\end{equation}
which coinsides with the critical peak intensity above which bright 
solitons does not exist in a focusing nonlocal medium when $\gamma<0$,
as discussed above.

It can be shown that the relation (\ref{critical}) represents the stability 
condition for the plane wave solution to equation (\ref{NLS1}) with a 
defocusing nonlinearity. 
For background intensities larger than $\rho_{\rm cr}$ the plane wave 
solutions becomes modulationally unstable \cite{wk-MI}.

\begin{figure}
  \vspace{0mm}
  \epsfig{file=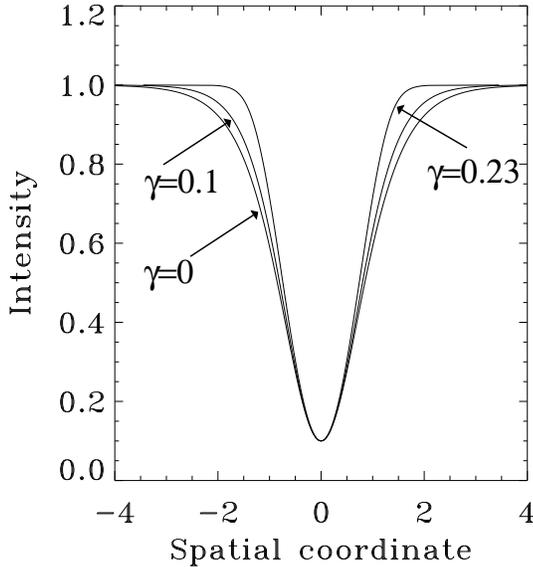,width=7cm,angle=0}
  \caption{Intensity profiles of dark solitons with $\rho_0=1$ and 
  $\rho_1=0.1$ for different values of the nonlocality parameter $\gamma$.}
  \label{fig5}
\end{figure}

\begin{figure}
  \epsfig{file=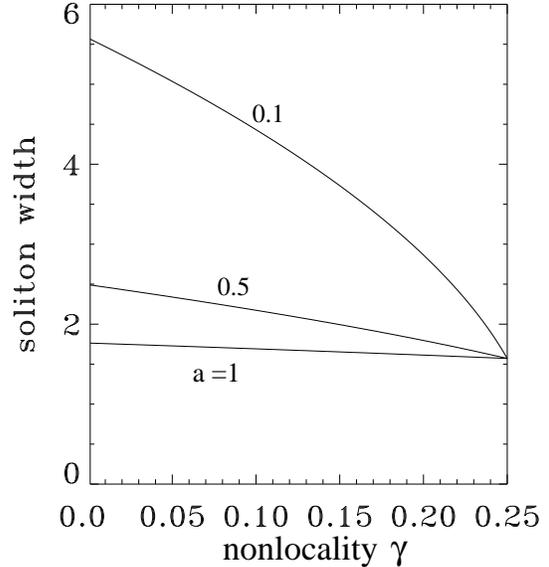,width=7cm,angle=0}
  \caption{Width of dark solitons versus degree of nonlocality $\gamma$ 
  for $\rho_0=1$ and different contrasts $a=(\rho_0-\rho_1)/\rho_0$.} 
  \label{fig6}
\end{figure}

Integrating Eq.~(\ref{dark-ODE}) once more we obtain
\begin{equation}
  \label{dark-amplitude}
  \pm \zeta = 
  \frac{1}{\delta_0}\tanh^{-1}\left(\frac{\delta}{\delta_0}\right)
  + \sqrt{4\gamma} \tan^{-1} (\sqrt{4\gamma}\delta), 
\end{equation}
which is an implicit relation between the soliton intensity $\rho$ and
the spatial coordinate $\zeta$, with the normalized intensity now given 
by $\delta^2(\rho)=(\rho-\rho_1)/(1-4\gamma\rho)$ and $\delta_0=\delta
(\rho_0)$.  For the soliton phase we obtain
\begin{equation}
  \label{dark-phase}
  \pm\phi = \tan^{-1}\left(\frac{\delta}{u_1}\right) - 
            u_1\sqrt{4\gamma}\tan^{-1} (\sqrt{4\gamma}\delta),
\end{equation}
where we have used the gauge invariance of Eq.~(\ref{nls1}) to remove 
a constant phase contribution (phase in the center). 
In Fig.~\ref{fig5} we present examples of the intensity profile of the 
dark solitons for different values of the nonlocality parameter $\gamma$.  
These graphs show that increasing nonlocality decreases the width of the 
soliton. 
To better illustrate this effect we plot in Fig.~\ref{fig6} the width
of the dark solitons (defined as the distance between points where the 
intensity is $(\rho_0+\rho_1)/2$) versus $\gamma$ for several values 
of the soliton contrast $a=(\rho_0-\rho_1)/\rho_0$. 
We see that unlike saturation, which tends to increase the width of dark 
solitons, nonlocality has the opposite effect.  
This is due to the fact that nonlocality causes the 
nonlinear index change to advance towards regions of lower light intensity.  
In the case of bright solitons this effect leads to soliton expansion.  
The plots in Fig.~\ref{fig6} also indicate the interesting effect that 
for strong nonlocality, all dark solitons acquire the same width.  
Actually, in this limit, i.e.~when $\gamma$ approaches the critical 
value $\gamma=1/(4\rho_0)$ the relation (\ref{dark-amplitude}) can 
be inverted leading to a simple expression for the soliton intensity 
profile
\begin{equation}
  \label{limit}
  \rho = \left \{
  \begin{array}{ll}
    \rho_0[1-a\cos^2(u_0\zeta)], & |u_0\zeta| \leq \pi/2 \\
    \rho_0,                      & |u_0\zeta| \geq \pi/2.
  \end{array} 
  \right.
\end{equation}
The solution (\ref{limit}) confirms that in this limit all dark solitons 
have the same width $\zeta_{\rm cr}=\pi/(2u_0)$ independent of the contrast.

\begin{figure}
  \epsfig{file=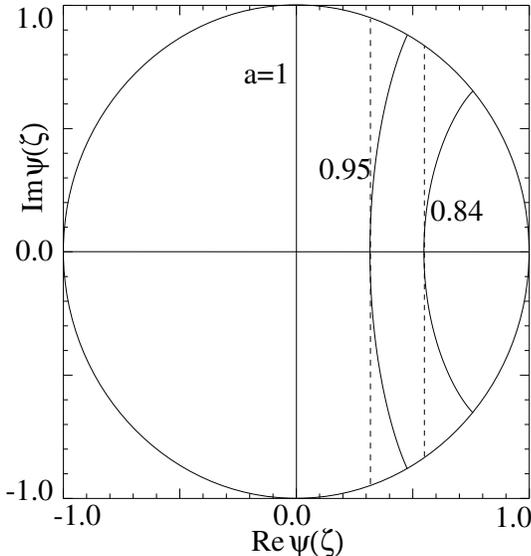,width=7cm,angle=0}
  \caption{Real and imaginary part of dark solitons with $\rho_0=1$ 
  in a weakly nonlocal medium with $\gamma=0.1$.} 
  \label{fig7}
\end{figure}

With a nontrivial phase structure, these dark solitons can be
represented in the complex plane describing the real and imaginary
part of the soliton amplitude $\psi(\zeta)=\sqrt{\rho}\exp(i\phi)$.
Any soliton solution is
then represented by a trajectory in this plane.  Figure \ref{fig7} shows
such a phase diagram ($\gamma=0.1$) for several values of the soliton
contrast. The circle represents the background intensity $\rho_0=1$ and 
nonlocal soliton solutions are plotted using solid lines.  The dashed line
corresponds to the same-contrast soliton solution in the purely local
case ($\gamma=0$).  The total phase change across the soliton is given
by the angle subtended by two lines connecting the origin with points of
intersection of the soliton curve and the circle.
One can see directly from Fig.~\ref{fig7} that the total phase change 
across the soliton in a nonlocal medium is always less than in the 
local case.  Furthermore, the soliton phase profile closely resembles 
that of a threshold nonlinearity \cite{darker}.

\begin{figure}
  \epsfig{file=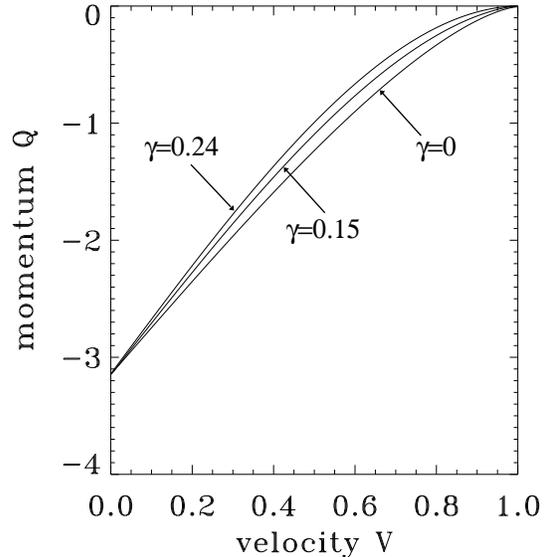,width=7cm,angle=0}
  \caption{Momentum $Q$ of dark solitons with $\rho_0=1$ versus 
  velocity $V$ for different degrees of nonlocality, $\gamma$.} 
  \label{fig8}
\end{figure}

The linear stability of dark solitons can be determined by considering
the conserved quantities of the nonlinear Schr\"odinger Eq.~(\ref{NLS1}) 
with defocusing nonlinearity, namely the renormalized Hamiltonian
\begin{equation}
  H = \frac{1}{2}\int_{-\infty}^\infty\left[|\partial_x \psi|^2+
      (|\psi|^2-\rho_0)^2 - \gamma(\partial_x|\psi|^2)^2\right]dx
\end{equation}
and the renormalized momentum
\begin{equation}
  Q = \frac{i}{2}\int_{-\infty}^\infty
  (\psi\partial_x\psi^*-\psi^*\partial_x\psi)
  (1-\frac{\rho_0}{|\psi|^2})dx.
\end{equation}
A soliton solution propagating with velocity $V$ corresponds to an 
extremum of the Hamiltonian for fixed momentum, $\delta(H-VQ)=0$.
It has been shown that the stability criterion for dark solitons is
determined by the dependence of the momentum on the velocity, $Q=Q(V)$ 
\cite{wk-yk}.  For dark solitons to be stable the derivative 
of the momentum with respect to the velocity must be positive,
\begin{equation}
  dQ/dV > 0.
\end{equation}
In the case of a nonlocal nonlinearity the expression for the
Momentum can be evaluated analytically to
\begin{eqnarray}
 Q & = & -2\rho_o\tan^{-1}\left(\frac{\delta_0}{\sqrt{\rho_1}}\right) 
         + (\rho_0-\rho_1)\left(\frac{\sqrt{\rho_1}}{\delta_0}\right)  
         \nonumber \\
   &   & + [1+4\gamma(2\rho_0-\rho_1)]
         \sqrt{\frac{\rho_1}{4\gamma}}\tan^{-1}(\sqrt{4\gamma}\delta_0)
\end{eqnarray}
The dependence of the momentum on the velocity of dark solitons with
background intensity $\rho_0=1$ is shown in Fig.~\ref{fig8} for several 
degrees of nonlocality $\gamma$. 
The monotonic increase of this function indicates that the dark 
solitons are stable.

\begin{figure}
  \epsfig{file=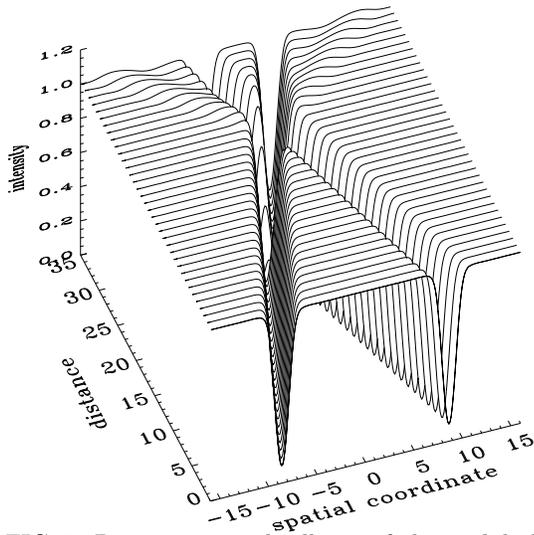,width=7cm,angle=0}
  \caption{Propagation and collision of identical dark solitons with
  $\rho_0=10\rho_1=1$ in a weakly nonlocal medium with $\gamma=0.1$}
  \label{fig9}
\end{figure}

This conclusion is confirmed by direct numerical simulations of 
Eq.~(\ref{NLS1}) with initial conditions given by the exact solution 
(\ref{dark-amplitude}-\ref{dark-phase}). 
Fig.~\ref{fig9} illustrates propagation and collision of two identical 
dark solitons with $\rho_0=10\rho_1=1$ in a weakly nonlocal medium
with $\gamma=0.1$.  While the solitons propagate in a stable fashion, 
their collision is clearly inelastic with radiation being emitted 
from the impact area.

In conclusion, we have studied the properties of bright and dark 
one-dimensional spatial solitons in a general weakly nonlocal 
Kerr-like medium, in which the change in refractive index due 
to local and nonlocal contributions are of the same sign.  
We have found an exact analytical solution for both bright and dark 
solitons and used it to find the existence and stability regions of 
both solutions. 

Stable bright soliton solutions were found to exist in focusing 
nonlocal media for all parameter values. 
Although they were previously predicted the solution was never 
explicitly written down. The effect of nonlocality is in some 
sense equivalent to that of saturation, - to smooth out the 
index profile and thereby increase the soliton width.

Previously unknown dark soliton solutions were found to exist in 
defocusing nonlocal media for background intensities below a certain 
critical value, which correspond to the intensity at which the 
plane wave solutions become unstable. 
We found that nonlocality leads to narrowing of the dark solitons,
with all dark solitons acquiring the same width in the strongly 
nonlocal limit, independent of the soliton contrast. 
The total phase change across the dark soliton was found to be less
that in the purely local case, with the phase profile resembling
that of a threshold nonlinearity.

Both nonlocal bright and dark spatial solitons appear to
be stable under propagation. Preliminary studies of soliton collision
revealed their inelastic character in analogy to collisions of
solitons of nonintegrable models.

O.~Bang acknowledges support from the Danish Technical Research Council
under Talent Grant No.~9800400.

\end{multicols}
\end{document}